\documentclass[preprint,superscriptaddress]{revtex4-1}

\usepackage{amsmath,amssymb,graphicx,subfigure}% Include figure files
\usepackage{dcolumn}% Align table columns on decimal point
\usepackage{bm}% bold math
\usepackage{soul}
\usepackage{xcolor}

\newcommand\be{\begin{equation}}
\newcommand\ee{\end{equation}}
\newcommand\bea{\begin{eqnarray}}
\newcommand\eea{\end{eqnarray}}

\newcommand\mpl{M_{\rm Pl}}

\definecolor{DarkBlue}{rgb}{0.15,0.15,0.85}
\usepackage[linktocpage=true]{hyperref}
\hypersetup{colorlinks=true, citecolor=DarkBlue, linkcolor=DarkBlue, urlcolor=DarkBlue}
\usepackage{hyperref}

\begin{document}

\title{The Upper Bound on the Tensor-to-Scalar Ratio Consistent with Quantum Gravity}

\author{Lina Wu}\email{wulina@xatu.edu.cn}
\affiliation{School of Sciences, Xi'an Technological University, Xi'an 710021, China}

\author{Qing Gao}\email{gaoqing1024@swu.edu.cn }
\affiliation{School of Physical Science and Technology,
Southwest University, Chongqing 400715, China}

\author{Yungui Gong}\email{yggong@hust.edu.cn}
\affiliation{School of Physics, Huazhong University of Science and Technology, \\Wuhan, Hubei 430074, China}

\author{Yiding Jia}
\email{jiayiding@itp.ac.cn}
\affiliation{CAS Key Laboratory of Theoretical Physics, Institute of Theoretical Physics, \\Chinese Academy of Sciences, Beijing 100190, China }
\affiliation{ School of Physical Sciences, University of Chinese Academy of Sciences, No.~19A Yuquan Road, Beijing 100049, China }

\author{Tianjun Li}\email{tli@itp.ac.cn}
\affiliation{CAS Key Laboratory of Theoretical Physics, Institute of Theoretical Physics, \\Chinese Academy of Sciences, Beijing 100190, China }
\affiliation{ School of Physical Sciences, University of Chinese Academy of Sciences, No.~19A Yuquan Road, Beijing 100049, China }

\begin{abstract}

We consider the polynomial inflation with the tensor-to-scalar ratio as large as possible which can be consistent with the Quantum Gravity (QG) corrections and Effective Field Theory (EFT). To get a minimal field excursion $\Delta\phi$ for enough e-folding number $N$, the inﬂaton field traverses an extremely flat part of the scalar potential, which results in the Lyth bound to be violated. We get a CMB signal consistent with Planck data by numerically computing the equation of motion for inflaton $\phi$ and using Mukhanov-Sasaki formalism for primordial spectrum. Inflation ends at Hubble slow-roll parameter $\epsilon_1^H=1$ or $\ddot{a}=0$. Interestingly, we find an excellent practical bound on the inflaton excursion in the format $a+b{\sqrt r}$, where $a$ is a tiny real number and $b$ is at the order 1.  To be consistent with QG/EFT and suppress the high-dimensional operators, we show that the concrete condition on inflaton excursion is $\frac{\Delta \phi}{M_{\rm Pl}} < 0.2 \times \sqrt{10}\simeq 0.632$. For $n_s=0.9649$, $N_e=55$, and  $\frac{\Delta \phi}{M_{\rm Pl}} < 0.632$, we predict that the tensor-to-scalar ratio is smaller than 0.0012 for such polynomial inflation to be consistent with QG/EFT.

%In addition, even for $n_s=0.9649$, $r = 0.0012$ and $N_e=55$, we show that $\phi/M_{\rm Pl} $ can be at the %order of $0.1$, and thus such polynomial inflation is indeed consistent with the QG corrections and EFT.

\end{abstract}


%\pacs{\dots}% PACS

\maketitle
%\vspace*{0.3cm}

%%%%%%%%%%%%%%%%%%%%%%%%%%%%%%%%%%%%%%%%%%%%%%%%%%%%%%%%%
%%%%%%%%%%%%%%%%%%%%%%%%%%%%%%%%%%%%%%%%%%%%%%%%%%%%%%%%%
\section{Introduction}

Inflation provides a natural solution to the well-known
flatness, horizon, and monopole problems, etc,
in the standard big bang cosmology~\cite{STAROBINSKY198099, Guth:1980zm, Linde:1981mu, Albrecht:1982wi}.
And the observed temperature fluctuations in the cosmic microwave background radiation (CMB)
strongly indicates an accelerated expansion at a very early stage of
our Universe evolution, {\it i.e.}, inflation. In addition,
the inflationary models predict the cosmological perturbations for the matter density and
spatial curvature due to the vacuum fluctuations of the inflaton, and thus can explain the primordial
power spectrum elegantly. Besides the scalar perturbation, the tensor perturbation is also generated.
Especially, it has special features in the B-mode of the CMB polarization data as a signature of the primordial inflation.

The Planck satellite measured the CMB temperature anisotropy with
an unprecedented accuracy. From the latest observational data~\cite{Akrami:2018odb,Aghanim:2018eyx}, the scalar spectral index $n_s$, the running of the scalar spectral index $\alpha_s \equiv d n_s/d\ln k$, the tensor-to-scalar ratio $r$, and the scalar amplitude $A_s$
for the power spectrum of the curvature perturbations  are respectively constrained to be
\begin{eqnarray}
&& n_s = 0.9649 \pm 0.0042(68\%~{\rm CL})~,~~\alpha_s=-0.0045\pm0.0067(68\%~{\rm CL})~,~~\nonumber\\
%-0.005\pm 0.0013(95\%~{\rm CL}) ~,~~\nonumber \\
&& r_{0.002} \le 0.056(95\%~{\rm CL})~,~~\ln[10^{10}A_s] = 3.044\pm{0.014}~.~\,
\label{Planck-A}
\end{eqnarray}
There is no sign of primordial non-Gaussianity in the 
CMB fluctuations. On the other hand, from the analysis of BICEP2/Keck CMB polarization experiments \cite{Ade:2014xna,Ade:2018gkx}, the upper limits are set to $r<0.07$ at $95\% ~\rm{CL}$. The future QUBIC experiment ~\cite{Battistelli:2010aa,Battistelli:2020siw} targets to constrain the tensor-to-scalar ratio of 0.01 at 95\% CL with two years of data. Therefore, the interesting question is how to construct the inflation models which can be consistent with the Planck results and have large tensor-to-scalar ratio. The energy scale of inflation is described by the ratio between the amplitude of the tensor mode and scalar mode of CMB \cite{Lyth:1984yz}
\begin{equation}
	V^{1/4}\simeq \left(\frac{r}{0.01}\right)^{1/4}\times10^{16}\rm ~GeV
\end{equation}
So, the present upper limit of $r<0.056$ gives the constraint $V^{1/4}< 1.5\times10^{16}\rm~GeV$, which is consistent with grand unified theories of fundamental interactions.

However, from the Lyth bound~\cite{Lyth:1996im}, the inflaton $\phi$ excursion $\Delta \phi$ is larger than about $ 2 M_{\rm Pl}$ for $r=0.01$ and $N \sim 55$, where $M_{\rm Pl}$ is the reduced Planck scale and $N$ is the number of e-folders. Thus, we will define the large tensor-to-scalar ratio as $r > 0.01$, and call the large field inflation if $\Delta \phi >  2 M_{\rm Pl}$. If the Lyth bound is valid, the big challenge to the inflationary models with large $r$ is the high-dimensional operators in the inflaton potential from the effective field theory (EFT) point of view, which may be generated by the Quantum Gravity (QG) effects, since such high-dimensional operators can not be suppressed by $M_{\rm Pl}$ and then are out of control. Three kinds of solutions to this the problem are: (1) The natural inflation since the quantum gravity corrections may be forbidden by the discrete symmetry~\cite{Freese:1990rb, Adams:1992bn}. However, the quantum gravity does not preserve the global symmetry.
(2) The phase inflation since the phase of a complex field may not play an important role in the non-renormalizable operators from quantum gravity effects~\cite{Li:2014vpa, McDonald:2014oza, McDonald:2014nqa}. (3) The polynomial inflation with the Lyth bound violation and small $\Delta \phi$~\cite{BenDayan:2009kv, Choudhury:2013iaa, Choudhury:2013iaa, Choudhury:2014kma, Antusch:2014cpa, Gao:2014pca}. Meanwhile, the (extended) Lyth bound or the bound of field excursion $\Delta\phi$ have been studies in different models~\cite{Efstathiou:2005tq,Baumann:2011ws,Garcia-Bellido:2014eva, Garcia-Bellido:2014wfa, Huang:2015xda, Linde:2016hbb, DiMarco:2017ihz}. However, no one has constructed a concrete, solid, and interesting inflationary model with $\Delta \phi \le {\cal O} (0.1)  M_{\rm Pl}$ so far. And also, the Gao-Gong-Li (GGL) bound on inflaton excursion~\cite{Gao:2014pca} is $\Delta\phi<0.1~\mpl$ with a modified tensor-to-scalar bound $r<0.02$. It's too low and can not be saturated, and it is not clear whether there exists a new practical bound. Note that, the minimal field excursion $\Delta\phi$ plays an important role to determine the upper bound of $r$.

In this paper, we shall study the polynomial inflation with the tensor-to-scalar ratio as large as possible which is consistent with the QG corrections and EFT. 
We do realize the small field inflation with a large $r$,
and then the Lyth bound is violated obviously. 
Interestingly, we find an excellent practical bound on the inflaton excursion
in the format $a+b{\sqrt r}$, where $a$ is a small real number and $b$ is at the order 1.
 To be consistent with QG/EFT and suppress the high-dimensional operators, we show that the concrete condition on inflaton excursion is $\frac{\Delta \phi}{M_{\rm Pl}} < 0.2 \times \sqrt{10}\simeq 0.632$. For $n_s=0.9649$, $N_e=55$, and  $\frac{\Delta \phi}{M_{\rm Pl}} < 0.632$, we predict that the tensor-to-scalar ratio is smaller than 0.0012 for such polynomial inflation to be consistent with QG/EFT.

%%%%%%%%%%%%%%%%%%%%%%%%%%%%%%%%%%%%%%%%%%%%%%%%%%%%%%%%%
%%%%%%%%%%%%%%%%%%%%%%%%%%%%%%%%%%%%%%%%%%%%%%%%%%%%%%%%%
\section{The polynomial inflation}
We will consider the order 5 polynomial inflation for numerical study, {\it i.e.},
\begin{equation}
V(\phi) = V_0\left[1+\sum_{m=1} \lambda_m (\phi-\phi_*)^m \right],~m=1, ~2, ~...,~5.
\end{equation}
The slow-roll parameters $\epsilon$, $\eta$, and $\xi^2$ are defined as
\begin{eqnarray}
\epsilon=\epsilon_1=\frac{M_{\rm Pl}^2(V^{\prime})^2}{2V^2}~,~
\eta=\epsilon_2=\frac{M_{\rm Pl}^2V^{\prime \prime}}{V}~,~
\xi^2=\epsilon_3=\frac{M_{\rm Pl}^4V^{\prime} V^{\prime \prime \prime}}{V^2}~,~
\end{eqnarray}
where $X' \equiv dX(\phi)/d\phi$. And the other two relevant slow-roll parameters \cite{Lyth:1998xn} in terms of the order 5 polynomial inflaton potential are
\begin{eqnarray}
\sigma^3=\epsilon_4=M^6_{\rm Pl}(V^{\prime})^2V^{\prime \prime \prime \prime}/V^3~,~
\delta^4(\phi)=\epsilon_5=M^8_{\rm Pl} (V^{\prime})^3 V^{\prime \prime \prime \prime \prime}/V^4~,~
\end{eqnarray}
The number of e-folding before the end of inflation is
\begin{equation}
\label{efolddef}
N(\phi)=\int_t^{t_e}Hdt\approx \frac{1}{M_{\rm Pl}}\int_{\phi_e}^\phi\frac{d\phi}{\sqrt{2\epsilon(\phi)}},
\end{equation}
where  the inflaton value $\phi_e$ at the end of inflation is determined by $\ddot{a}=0$ or $\epsilon=1$, where $a$ is the scale factor.  If $\epsilon(\phi)$ is a monotonic function during inflation, we have $\epsilon(\phi)>\epsilon(\phi_*)=\epsilon=r/16$, and then get the Lyth bound \cite{Lyth:1996im}
\begin{eqnarray}
\frac{\Delta \phi}{\mpl} \equiv |\phi_*-\phi_e| > \sqrt{2\epsilon}~ N_e =\sqrt{r/8}~ N_e ~,
\end{eqnarray}
where the subscript $``*"$ means the value at the horizon crossing, and $n_s$, $\alpha_s$ and $r$ are evaluated at $\phi_*$. For example, the Lyth bound gives $\Delta \phi = 1.947 M_{\rm Pl}$ for $r=0.01$ and $N_e =55$. Thus, to realize the small field inflation with large $r$, the Lyth bound must be violated. In other words, $\epsilon(\phi)$ should not be a monotonic function and has at least one minimum between $\phi_*$ and $\phi_e$~\cite{BenDayan:2009kv}.

%%%%%%%%%%%%%%%%%%%%%%%%%%%%%%%%%%%%%%%%%%
\subsection{Numerical results}
To compute observable quantities for the CMB, we numerically evolve the scalar field according to the Friedman equation and equation of motion for $\phi$:
\begin{equation}
\begin{split}
&	\ddot{\phi}+3H\dot{\phi}+V'=0,\\
&	H^2\equiv\left(\frac{\dot{a}}{a}\right)^2=\frac{8\pi G}{3}\left(\frac{1}{2}\dot{\phi}+V\right).\\
\end{split}\label{eq:field}
\end{equation}
For numerical purposes it is more convenient to rewrite the inflaton evolution as a function of conformal time $\tau$ rather than time $t$. Using $\tau=\frac{dt}{a}$ the cosmological evolution equation becomes 
\begin{eqnarray}
\frac{d^2\phi}{d\tau^2}+2aH\dfrac{d\phi}{d\tau}+a^2V'=0~.~\,
\end{eqnarray}
For convenience, we use the slow-roll parameters defined via Hubble parameter: $$\epsilon^H_1=-\frac{\dot{H}}{H^2}\approx\epsilon~,~\epsilon^H_{2}=\frac{\dot{\epsilon^H_1}}{H\epsilon^H_1}\approx4\epsilon-2\eta~,~\epsilon^H_{i+1}=\frac{\dot{\epsilon^H_i}}{H\epsilon^H_i}~(i=2,3,4),$$ 
where dots denote derivatives respect to the cosmic time. The inflation ends at $\ddot{a}=0$ or $\epsilon_1^H=1$.

To get the minimal field excursion $\Delta\phi$ and enough e-folding number $N$ for each $r$, the inﬂaton field must traverse an extremely flat part of the scalar potential. This is similar to ultra-slow-roll inflation (USR)~\cite{PhysRevD.69.084005,PhysRevD.72.023515,2016PhRvD..94j3515H,DIMOPOULOS2017262,Easther:2006qu}. There are 3 inflection points for the 5-th degree polynomial. We find a set of parameters which make sure there is an inflection point $\phi_{infl}$ during inflation and the derivative of $V(\phi)$ at the inflection point should be small. In such situation, an exact scalar spectral index $n_s$ and tensor-to-scalar ratio $r$ will numerically calculated by primordial spectrum using the Mukhanov-Sasaki formalism \cite{10.1143/PTP.76.1036,Mukhanov:1988jd}. The scalar mode $u_k=-z\mathcal{R}$ and tensor mode $v_k$ of primordial perturbation are given by 
\begin{eqnarray}
\frac{d^2u_k}{d\tau^2}+\left(k^2-\frac{1}{z}\frac{d^2z}{d\tau^2}\right)u_k&=&0,\\
\frac{d^2v_k}{d\tau^2}+\left(k^2-\frac{1}{a}\frac{d^2a}{d\tau^2}\right)v_k&=&0,
\end{eqnarray}
where $\mathcal{R}$ is the comoving curvature perturbation,  and $z=\frac{a}{\mathcal{H}}\frac{d\phi}{d\tau}$. In the limit $k\to\infty$, the modes are in the Bunch-Davies vacuum $u_k\to e^{-ik\tau}/\sqrt{2k}$ and $v_k\to e^{-ik\tau}/\sqrt{2k}$. The scalar and tensor spectrum of primordial perturbations at CMB scales can be accurately expressed as
\begin{eqnarray}
\mathcal{P_R}(k)&=A_s\left(\frac{k}{k_*}\right)^{n_s-1+\frac{\alpha_s}{2}\ln \frac{k}{k_*}+\cdots},~~
\mathcal{P}_t(k)&=A_t\left(\frac{k}{k_*}\right)^{n_t+\cdots}.
\end{eqnarray}
Then the spectral index, running of the spectral index and tensor-to-scalar ratio at $k_*=aH=0.05~\rm MPc^{-1}$ can determined by
\begin{equation}
n_s-1=\frac{d\ln\mathcal{P_R}}{d\ln k}~,~ r=\dfrac{\mathcal{P}_T}{\mathcal{P_R}}.\label{eq:nsr_num}
\end{equation}

\begin{figure}[!h]
	\includegraphics[width=0.4\linewidth]{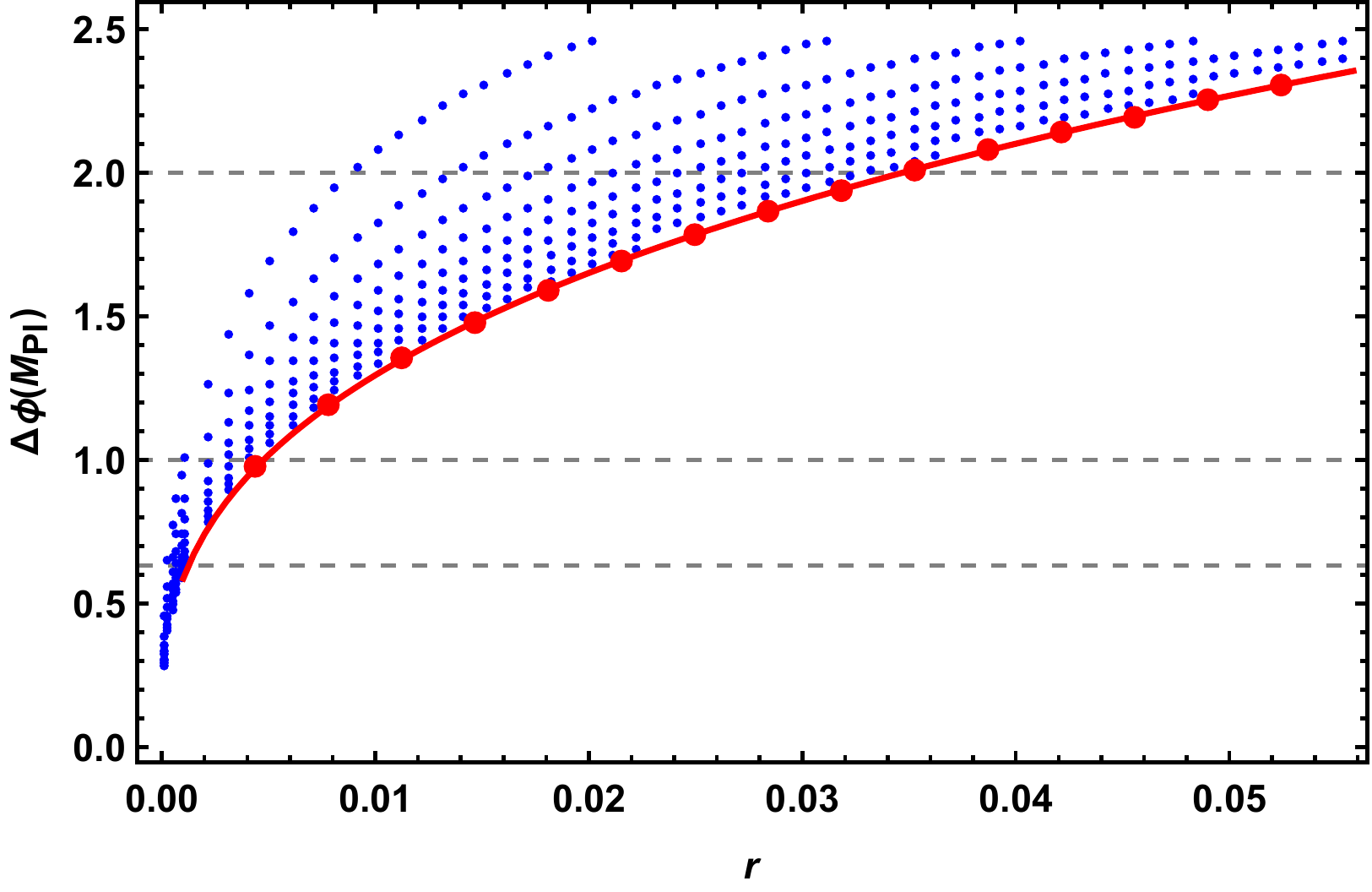}
	\caption{$r$ vs $\Delta\phi$ for the polynomial inflation. Here the e-folding number, the scalar spectral index and the relative running are fixed to be $N=55$, $n_s=0.9649$, and $\alpha_s=-0.0045$, respectively. The red line corresponds to the the low bound on the inflaton excursion. The three horizontal dashed lines  correspond to $\Delta\phi=0.632,~1.0,~2.0~\mpl$, respectively.}\label{fig:r_deltphi_slow}
\end{figure}

To simplify the numerical study, we choose $N_e = 55$, as well as the best fit for $n_s$ and $\alpha_s$, {\it i.e.}, $n_s = 0.9649$, $\alpha_s = -0.0045$ and $A_s=2.20\times10^{-9}$ \cite{Ade:2014xna,Ade:2018gkx}. Without loss of generality, we will take $\phi_*=0$. $r$ and $\Delta\phi$ for the 5-th degree polynomial inflation are given in Fig.~\ref{fig:r_deltphi_slow}, where the red point-line is corresponding to the low bound on inflaton excursion. The inflation ends at $\epsilon_1^H=1$ or $\ddot{a}=0$. 
\begin{table}[!h]
	\begin{tabular}{m{0.1\textwidth}<{\centering}m{0.12\textwidth}<{\centering}m{0.12\textwidth}<{\centering}m{0.12\textwidth}<{\centering}m{0.12\textwidth}<{\centering}m{0.12\textwidth}<{\centering}<{\centering}m{0.12\textwidth}<{\centering}}
		\hline\hline
		$r$  & $\lambda_1(10^{-2})$& $\lambda_2(10^{-3})$& $\lambda_3(10^{-3})$ & $\lambda_4(10^{-2})$ & $\lambda_5(10^{-2})$&$\Delta\phi(\mpl)$\\
		\hline
		0.01&-3.6188& -10.0437& -9.8543& 32.1351& -35.5515&1.313\\
		0.02&-5.1138& -9.03945&-6.64997& 16.0928&-12.7293&1.673 \\
		0.03&-6.25887& -8.08913& -5.19385& 10.7428& -7.0106&1.928 \\
		0.04&-7.22116& -7.06825& -4.32425& 8.07043& -4.61105&2.129 \\
		0.05&-8.05932& -6.08622& -3.73725& 6.47909& -3.34879& 2.297\\
		0.056&-8.51818& -5.47844& -3.4711& 5.79987& -2.85268&2.385 \\
		\hline
		0.0012&-1.25467& -11.007& -29.7032& 267.333& -844.752&0.632\\
		0.0046&-2.45598& -10.6889& -1.49114& 69.7688&-113.004&1.000\\
		0.0335&-7.5746& -5.6969& -4.1581& 7.1323& -3.8853&2.000\\
		\hline\hline
	\end{tabular}
	\caption{The low bound of inflaton excursion and the parameters for  inflation potential. Here e-folding number, the scalar spectral index and the relative running are fixed to the central value, i.e. $N=55$, $n_s=0.9649$, and $\alpha_s=-0.0045$.  }\label{tab:e0.01}
\end{table}

\begin{figure}[!h]
	\includegraphics[width=0.4\linewidth]{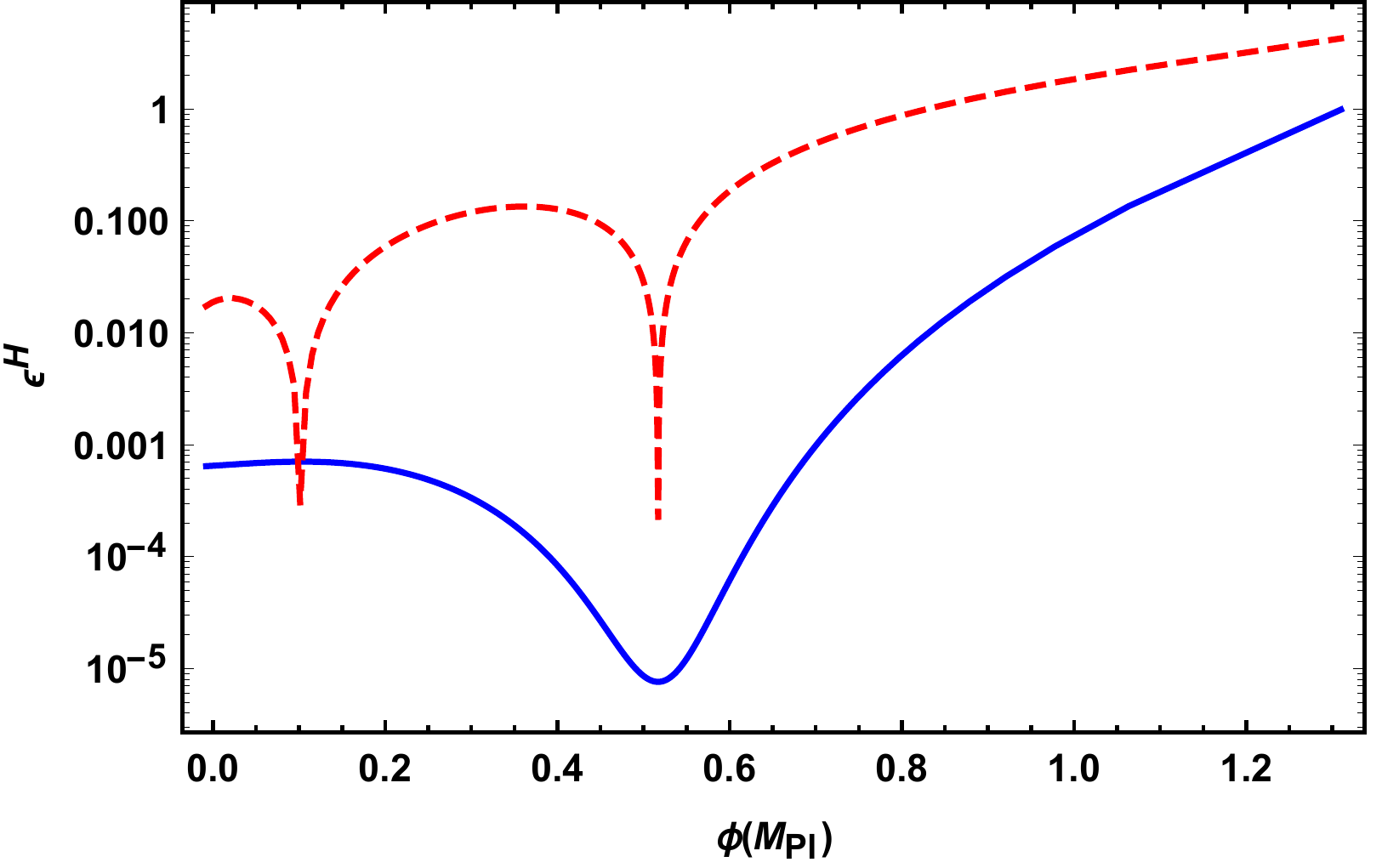}
	\caption{The evolution of Hubble flow slow-roll parameters $\epsilon_1^H$ (blue solid line) and $\epsilon_2^H$ (red dashed line).}\label{fig:slowroll_H}
\end{figure}

\begin{figure}[!h]
	\subfigure[]{\includegraphics[width=0.4\linewidth]{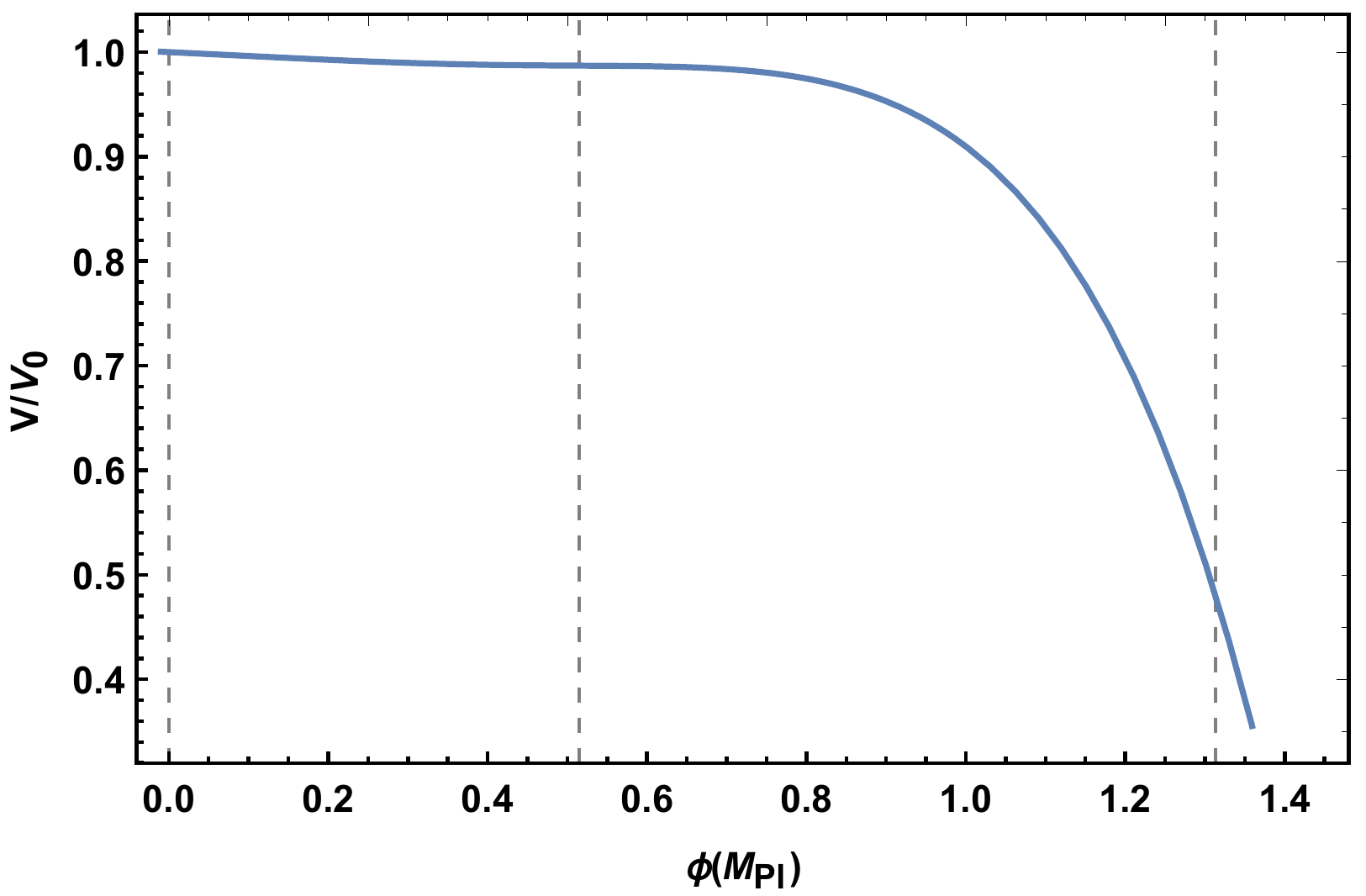}}~~
	\subfigure[]{\includegraphics[width=0.4\linewidth]{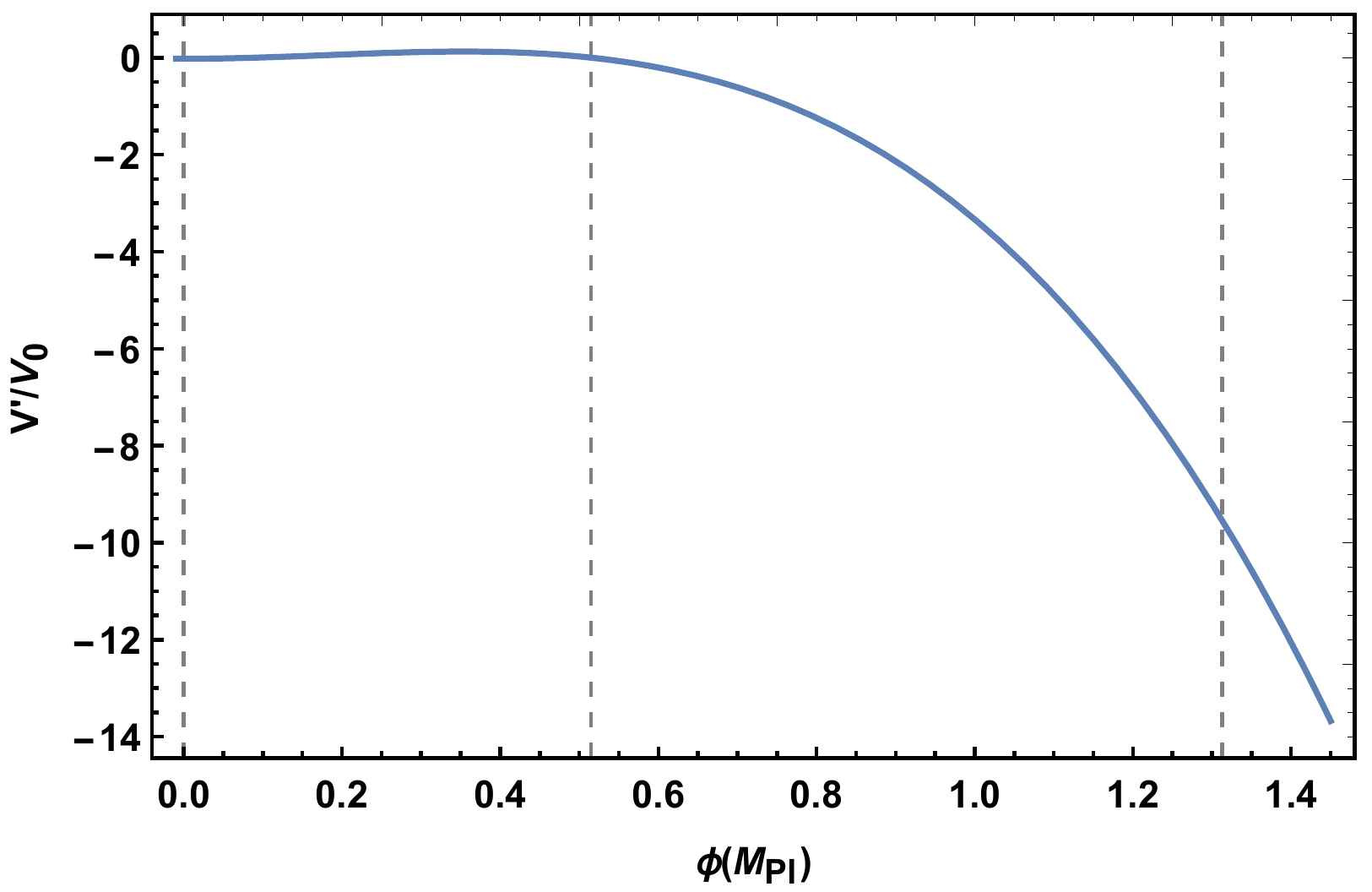}}
	\caption{The potential $V/V_0$ and its first derivative $V'/V_0$ with the parameters $\lambda_i=(-3.6188\times10^{-2}, -10.0437\times10^{-3},-9.8543\times10^{-3}, 32.1351\times10^{-2}, -35.5515\times10^{-2})$ . The vertical dashed lines correspond to the horizon crossing point $\phi_*=0$, the inflection point $\phi_{infl}=0.516~\mpl$ and the end of inflation $\phi_e=1.313~\mpl$.}\label{Fig:pot}
\end{figure}

To be concrete, we present some examples. Taking several tensor-to-scalar ratios $r=0.01 \sim0.056$, we try our best to get the minimal inflation excursions numerically. The results are shown in Tab. \ref{tab:e0.01}. To understand why our polynomial potential violates the Lyth bound but is consistent with the Planck results, we plot the hubble slow-roll parameters $\epsilon_1^H$ and $\epsilon_2^H$ in Fig.~\ref{fig:slowroll_H} for the model with $r=0.01$. The potential $V/V_0$ and its first derivative $V'/V_0$ with the parameters $\lambda_{i}=(-3.6188\times10^{-2}, -10.0437\times10^{-3},-9.8543\times10^{-3}, 32.1351\times10^{-2}, -35.5515\times10^{-2})$ are shown in Fig.~\ref{Fig:pot}. There is an inflection point $\phi_{infl}=0.515~\mpl$, and $V'/V_0(\phi_{infl})$ is close to zero. The evolution of inflaton is extremely slow around $\phi_{infl}$ and the slow-roll parameters $\epsilon_1^H$ decreases several orders of magnitude at $\phi_{infl}$.  This means that the inflation enters the USR.


%%%%%%%%%%%%%%%%%%%%%%%%%%%%%%%%%%%%%%%%%%%%%%%%%%%%%%%%%
\subsection{Slow roll approximation}
In the slow-roll inflation with inflaton potential $V(\phi)$, the observations are 
\begin{equation}
\begin{split}
n_{s}  =  1+2\eta-6\epsilon~,~
r =  16\epsilon~,~
\alpha_s  = 16\epsilon\eta-24\epsilon^2-2\xi^2~,
A_s=\dfrac{V}{24\pi^2\epsilon}~,~
\end{split}\label{eq:nsr}
\end{equation}
Since the inflation potential has order 5 polynomial, we need take into account the higher order corrections for $n_s$ and $r$. The higher order corrections \cite{STEWART1993171,Lyth:1998xn,CASADIO20051} are given in 
\begin{equation}
\begin{split}
\Delta n_{s} & = 2\left[\frac{1}{3}\eta^{2}-\left(\frac{5}{3}+12C\right)\epsilon^{2}+(8C-1)\epsilon\eta-(C-\frac{1}{3})\xi^2\right]~,\\
\Delta r& =  \frac{32\epsilon}{3}(3C-1)(2\epsilon-\eta)~,\\
\end{split}\label{eq:high_order}
\end{equation}
where $C=-2+\ln2+\gamma\simeq-0.7296$ with $\gamma$ the Euler\textendash Mascheroni constant. 
The slow-roll parameters at the horizon crossing $\phi_*=0$ are approximated as
\begin{equation}
\label{polyslow1}
\epsilon=\frac{\lambda_1^2}{2}~,~~ \eta=2\lambda_2~,~~
\xi^2=6\lambda_1\lambda_3.
\end{equation}
Then, the parameters $V_0$ and $\lambda_{1,2,3}$ can be determined by the observations in Eq.(\ref{eq:nsr}) as follows: 
\begin{equation}
\begin{split}
V_0=\frac{3}{2}\pi^2P_s r^{(1)}&~~,~~\lambda_1 =-\sqrt{r^{(1)}/8}~,~\\
\lambda_2 = \frac{n_s^{(1)}-1+3\lambda_1^2}{4}&~~,~~
\lambda_3 =-\frac{\alpha_s+6\lambda_1^4-16\lambda_1^2\lambda_2}{12\lambda_1}.  
\end{split}\label{eq:obs_lam}
\end{equation}
Therefore, we can get the higher order corrections of these observations,
\begin{equation}
\begin{split}
\Delta n_s  &=\frac{1}{384}\left[64\left(\alpha_s(6C-2)+(n_s^{(1)}-1)^2\right)+88r^{(1)}(n_s^{(1)}-1)+7(r^{(1)})^2\right]~,\\
\Delta r &=-\frac{r^{(1)}}{3}(3C-1)\left(n_s^{(1)}-1+\frac{r^{(1)}}{8}\right)~.\\
\end{split}
\end{equation}
Solving these equations with $\alpha_s=-0.0045$, we find that the second-order corrections will give a tiny contribution $\Delta r<-1.5\times 10^{-3}$ for $r<0.056$ and $\Delta n_{s}\sim4.9\times10^{-3}$ for $n_s=0.9649$. These corrections for CMB are still within $2~\sigma$ range of Planck data. %And the corrections $\Delta r$or $\Delta n_s$ decreases as $r$ increases. So, in the following calculations, we can first determined the parameters $\lambda_{1,2,3}$ by the observations. 
%%%%%%%%%%%%%%%%%%%%%%%%%%%%%%%%%%%%%%%%%%%%%%%%%%%%%%%%%

\subsection{The Lower Bound on Inflaton Excursion}
\begin{figure}[!h]
	\centering
	\subfigure[]{\includegraphics[width=0.4\linewidth]{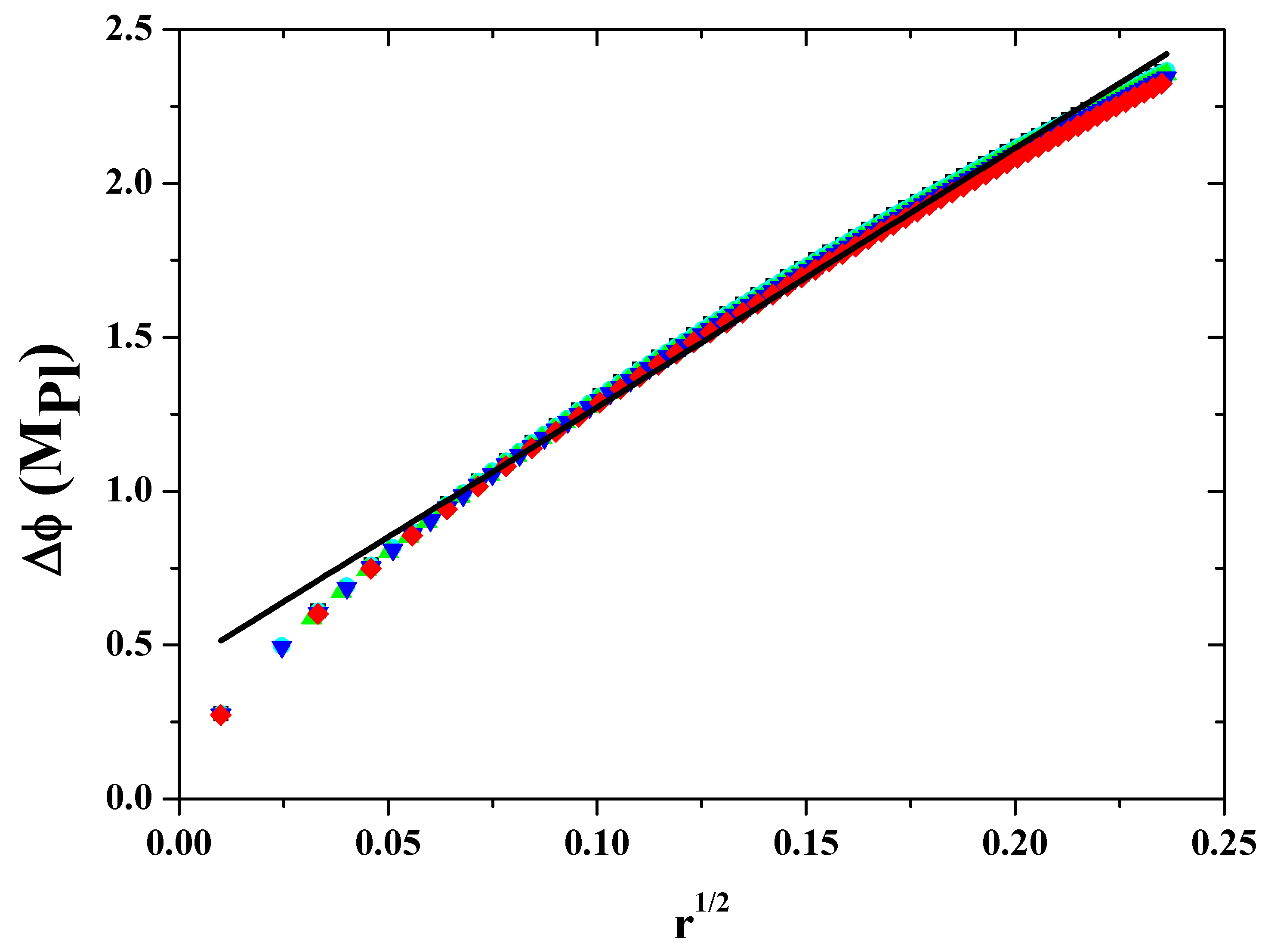}}
	\subfigure[]{\includegraphics[width=0.4\linewidth]{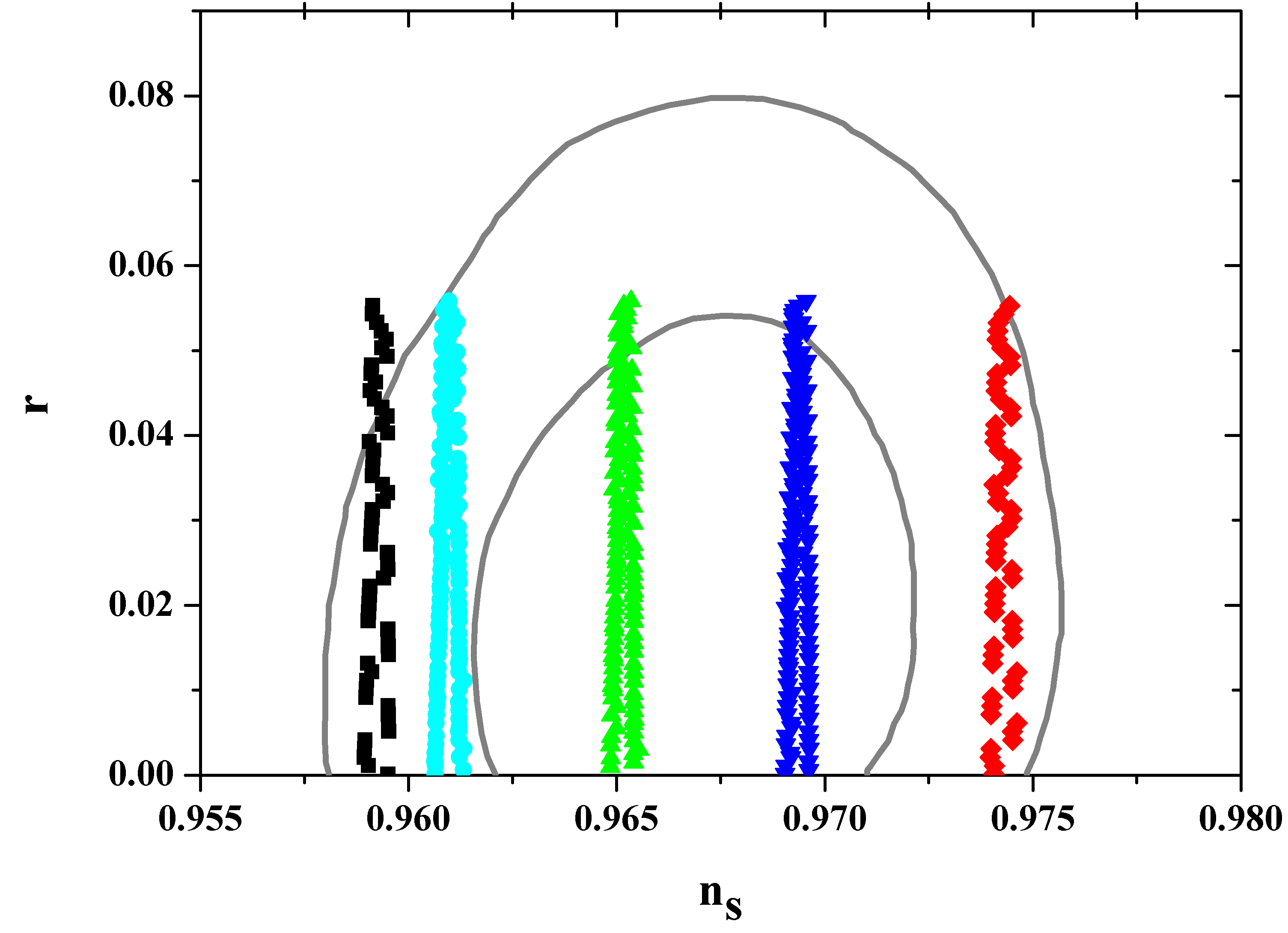}}
	\caption {The low bounds on inflaton excursions. The black line  corresponds to the fitted linear equation $a+b\sqrt{r}$. The different color points correspond to various central $n_s$. The range of $r$ is $r<0.056$.}\label{Fig:fit}
\end{figure}

For slow-roll inflation, we obtain $\Delta \phi/\mpl =  b \sqrt{r}$ if $\epsilon(\phi)$ is 
a constant during inflation. However, in general, $\epsilon(\phi)$ is a $\phi$ dependent varying function during inflation. Interestingly, for the slow-roll inflation with polynomial inflaton potential, 
we find that the generic lower bound on inflaton excursion is a linear function of $\sqrt{r}$, {\it i.e.}, $\Delta \phi/\mpl = a + b \sqrt{r}$ approximately. The results for a fixed $\alpha_s=-0.0042$ and $N=55$ are show in Fig.~\ref{Fig:fit}. The black line in Fig.~\ref{Fig:fit}(a) is the fitted linear equation $0.4569 + 8.2247 \sqrt{r}$ for fixed $n_s=0.9649$. We also check the field excursion variation as $n_s$ increases. The different color points in Fig.~\ref{Fig:fit} are corresponding to different central value of $n_s$. For the variation of $n_s$, the low bound of $\Delta\phi$ is pushed toward to a smaller value, which is consistent with previous results~\cite{DiMarco:2017ihz}. On the other hand, after fixing the tensor-to-scalar ratio $r=0.01$, we compare the field excursion results from Eq.~(\ref{eq:field}) and slow-roll approximation in Eq.~(\ref{eq:high_order}) for $n_s=0.9625,~0.9655,~0.9685$. The results are concluded in Tab.~\ref{tab:ns_deltaphi}. We can find that the field excursion become smaller at order $10^{-3}$ as $n_s$ increases.

\begin{table}
	\begin{tabular}{m{0.1\textwidth}<{\centering}m{0.15\textwidth}<{\centering}m{0.15\textwidth}<{\centering}}
		\hline\hline
		$n_s$&$\Delta\phi(\mpl)$&$\delta\phi(\mpl)$\\\hline
		0.9625&1.30117&1.11196\\
		0.9655&1.29696&1.10843\\
		0.9685&1.29244&1.10351\\
		\hline\hline
	\end{tabular}
	\caption{The low bound of inflaton excursion. $\Delta\phi$ is numerically calculated from Eq.~(\ref{eq:field}), and $\delta\phi$ is calculated with high order corrections under the slow roll approximation. Here the tensor-to-scalar ratio is fixed to be $r=0.01$.}\label{tab:ns_deltaphi}
\end{table}

%%%%%%%%%%%%%%%%%%%%%%%%%%%%%%%%%%%%%%%%%%%%%%%%%%%%%%%%%
%%%%%%%%%%%%%%%%%%%%%%%%%%%%%%%%%%%%%%%%%%%%%%%%%%%%%%%%%
\subsection{ The Consistency with QG and EFT}
To be consistent with QG/EFT and suppress the high-dimensional operators, we require that $\frac{\phi}{M_{\rm Pl}}$ should be at the order of $0.1$, {\it i.e.}, $\frac{0.1}{\sqrt{10}} \le \frac{\phi}{M_{\rm Pl}} \le 0.1\times \sqrt{10}$. To minimize the absolute value of $\frac{\phi}{M_{\rm Pl}}$, we can choose $-\phi_i=\phi_e=\frac{\Delta \phi}{2}$, and then $\frac{\phi}{M_{\rm Pl}} \le \frac{\Delta \phi}{2M_{\rm Pl}}$. Thus, to be consistent with QG and EFT, we require $\frac{\Delta \phi}{M_{\rm Pl}} < 0.2 \times \sqrt{10}\simeq 0.632$.

With QG/EFT effects, inflation excursions are smaller than $\Delta\phi<0.632\mpl$. We predict that the upper bound on tensor-to-scalar ratio $r$ for single field slow-roll inflation is $r\le 0.0012$. The corresponding parameters are also shown in Tab. \ref{tab:e0.01}. Meanwhile, for the inflaton excursions $\Delta\phi\sim1 M_{Pl}$ and $\Delta\phi\sim2 M_{Pl}$, we find that the maximal tensor-to-scalar ratios are $r \sim 0.0046$ and $r \sim 0.0335$, respectively for $N_e=55$. While the Lyth bound gives $r \sim 0.0026$ and $r \sim 0.0106$, respectively.
Thus, the Lyth bound is indeed violated in our polynomial inflation. Our results are consistent with the current cosmological observation data and might be probed by the future observed results of CMB polarization missions~\cite{Kogut:2011xw,Bouchet:2011ck,Matsumura:2013aja,lazear2014primordial,essinger2014class,Finelli:2016cyd} and gravitational waves experiments~\cite{Crowder:2005nr}, which will detect the tensor-to-scalar ratio at the order of $10^{-3}$.

%%%%%%%%%%%%%%%%%%%%%%%%%%%%%%%%%%%%%%%%%%%%%%%%%%%%%%%%%
%%%%%%%%%%%%%%%%%%%%%%%%%%%%%%%%%%%%%%%%%%%%%%%%%%%%%%%%%

\section{Conclusions}
We have considered the polynomial inflation with the tensor-to-scalar ratio as large as possible which is consistent with the QG corrections and EFT. We got the small field inflation with large $r$, and then the Lyth bound is violated obviously, since the evolution of the slow-roll parameters are very slow and the inflation enters ultra-slow-roll around the inflection point. Interestingly, we found an excellent practical bound on the inflaton excursion in the format $a+b{\sqrt r}$, where $a$ is a small real number and $b$ is at the order 1.  To be consistent with QG/EFT and suppress the high-dimensional operators, we show that the concrete condition on inflaton excursion is $\frac{\Delta \phi}{M_{\rm Pl}} < 0.2 \times \sqrt{10}\simeq 0.632$. For $n_s=0.9649$, $N_e=55$, and  $\frac{\Delta \phi}{M_{\rm Pl}} < 0.632$, we predict that the tensor-to-scalar ratio is smaller than 0.0012 for such polynomial inflation to be consistent with QG/EFT.

\section*{Acknowledgments}
This work was supported in part by the Projects 11875062, 11875136, and 11947302 supported by the National Natural Science Foundation of China, by the Major Program of the National Natural Science Foundation of China under Grant No. 11690021, by the Key Research Program of Frontier Science, CAS, and by the Scientific Research Program 2020JQ-804 supported by Natural Science Basic Research Plan in Shanxi Province of China.

\bibliographystyle{apsrev4-1}
\bibliography{SFILr}

\begin{thebibliography}{46}%
\makeatletter
\providecommand \@ifxundefined [1]{%
 \@ifx{#1\undefined}
}%
\providecommand \@ifnum [1]{%
 \ifnum #1\expandafter \@firstoftwo
 \else \expandafter \@secondoftwo
 \fi
}%
\providecommand \@ifx [1]{%
 \ifx #1\expandafter \@firstoftwo
 \else \expandafter \@secondoftwo
 \fi
}%
\providecommand \natexlab [1]{#1}%
\providecommand \enquote  [1]{``#1''}%
\providecommand \bibnamefont  [1]{#1}%
\providecommand \bibfnamefont [1]{#1}%
\providecommand \citenamefont [1]{#1}%
\providecommand \href@noop [0]{\@secondoftwo}%
\providecommand \href [0]{\begingroup \@sanitize@url \@href}%
\providecommand \@href[1]{\@@startlink{#1}\@@href}%
\providecommand \@@href[1]{\endgroup#1\@@endlink}%
\providecommand \@sanitize@url [0]{\catcode `\\12\catcode `\$12\catcode
  `\&12\catcode `\#12\catcode `\^12\catcode `\_12\catcode `\%12\relax}%
\providecommand \@@startlink[1]{}%
\providecommand \@@endlink[0]{}%
\providecommand \url  [0]{\begingroup\@sanitize@url \@url }%
\providecommand \@url [1]{\endgroup\@href {#1}{\urlprefix }}%
\providecommand \urlprefix  [0]{URL }%
\providecommand \Eprint [0]{\href }%
\providecommand \doibase [0]{http://dx.doi.org/}%
\providecommand \selectlanguage [0]{\@gobble}%
\providecommand \bibinfo  [0]{\@secondoftwo}%
\providecommand \bibfield  [0]{\@secondoftwo}%
\providecommand \translation [1]{[#1]}%
\providecommand \BibitemOpen [0]{}%
\providecommand \bibitemStop [0]{}%
\providecommand \bibitemNoStop [0]{.\EOS\space}%
\providecommand \EOS [0]{\spacefactor3000\relax}%
\providecommand \BibitemShut  [1]{\csname bibitem#1\endcsname}%
\let\auto@bib@innerbib\@empty
%</preamble>
\bibitem [{\citenamefont {Starobinsky}(1980)}]{STAROBINSKY198099}%
  \BibitemOpen
  \bibfield  {author} {\bibinfo {author} {\bibfnamefont {A.}~\bibnamefont
  {Starobinsky}},\ }\href {\doibase
  https://doi.org/10.1016/0370-2693(80)90670-X} {\bibfield  {journal} {\bibinfo
   {journal} {Physics Letters B}\ }\textbf {\bibinfo {volume} {91}},\ \bibinfo
  {pages} {99 } (\bibinfo {year} {1980})}\BibitemShut {NoStop}%
\bibitem [{\citenamefont {Guth}(1987)}]{Guth:1980zm}%
  \BibitemOpen
  \bibfield  {author} {\bibinfo {author} {\bibfnamefont {A.~H.}\ \bibnamefont
  {Guth}},\ }\href {\doibase 10.1103/PhysRevD.23.347} {\bibfield  {journal}
  {\bibinfo  {journal} {Adv. Ser. Astrophys. Cosmol.}\ }\textbf {\bibinfo
  {volume} {3}},\ \bibinfo {pages} {139} (\bibinfo {year} {1987})}\BibitemShut
  {NoStop}%
\bibitem [{\citenamefont {Linde}(1987)}]{Linde:1981mu}%
  \BibitemOpen
  \bibfield  {author} {\bibinfo {author} {\bibfnamefont {A.~D.}\ \bibnamefont
  {Linde}},\ }\href {\doibase 10.1016/0370-2693(82)91219-9} {\bibfield
  {journal} {\bibinfo  {journal} {Adv. Ser. Astrophys. Cosmol.}\ }\textbf
  {\bibinfo {volume} {3}},\ \bibinfo {pages} {149} (\bibinfo {year}
  {1987})}\BibitemShut {NoStop}%
\bibitem [{\citenamefont {Albrecht}\ and\ \citenamefont
  {Steinhardt}(1987)}]{Albrecht:1982wi}%
  \BibitemOpen
  \bibfield  {author} {\bibinfo {author} {\bibfnamefont {A.}~\bibnamefont
  {Albrecht}}\ and\ \bibinfo {author} {\bibfnamefont {P.~J.}\ \bibnamefont
  {Steinhardt}},\ }\href {\doibase 10.1103/PhysRevLett.48.1220} {\bibfield
  {journal} {\bibinfo  {journal} {Adv. Ser. Astrophys. Cosmol.}\ }\textbf
  {\bibinfo {volume} {3}},\ \bibinfo {pages} {158} (\bibinfo {year}
  {1987})}\BibitemShut {NoStop}%
\bibitem [{\citenamefont {Akrami}\ \emph {et~al.}(2018)\citenamefont {Akrami}
  \emph {et~al.}}]{Akrami:2018odb}%
  \BibitemOpen
  \bibfield  {author} {\bibinfo {author} {\bibfnamefont {Y.}~\bibnamefont
  {Akrami}} \emph {et~al.} (\bibinfo {collaboration} {Planck}),\ }\href@noop {}
  {\  (\bibinfo {year} {2018})},\ \Eprint {http://arxiv.org/abs/1807.06211}
  {arXiv:1807.06211 [astro-ph.CO]} \BibitemShut {NoStop}%
\bibitem [{\citenamefont {Aghanim}\ \emph {et~al.}(2018)\citenamefont {Aghanim}
  \emph {et~al.}}]{Aghanim:2018eyx}%
  \BibitemOpen
  \bibfield  {author} {\bibinfo {author} {\bibfnamefont {N.}~\bibnamefont
  {Aghanim}} \emph {et~al.} (\bibinfo {collaboration} {Planck}),\ }\href@noop
  {} {\  (\bibinfo {year} {2018})},\ \Eprint {http://arxiv.org/abs/1807.06209}
  {arXiv:1807.06209 [astro-ph.CO]} \BibitemShut {NoStop}%
\bibitem [{\citenamefont {Ade}\ \emph {et~al.}(2014)\citenamefont {Ade} \emph
  {et~al.}}]{Ade:2014xna}%
  \BibitemOpen
  \bibfield  {author} {\bibinfo {author} {\bibfnamefont {P.}~\bibnamefont
  {Ade}} \emph {et~al.} (\bibinfo {collaboration} {BICEP2}),\ }\href {\doibase
  10.1103/PhysRevLett.112.241101} {\bibfield  {journal} {\bibinfo  {journal}
  {Phys. Rev. Lett.}\ }\textbf {\bibinfo {volume} {112}},\ \bibinfo {pages}
  {241101} (\bibinfo {year} {2014})},\ \Eprint {http://arxiv.org/abs/1403.3985}
  {arXiv:1403.3985 [astro-ph.CO]} \BibitemShut {NoStop}%
\bibitem [{\citenamefont {Ade}\ \emph {et~al.}(2018)\citenamefont {Ade} \emph
  {et~al.}}]{Ade:2018gkx}%
  \BibitemOpen
  \bibfield  {author} {\bibinfo {author} {\bibfnamefont {P.}~\bibnamefont
  {Ade}} \emph {et~al.} (\bibinfo {collaboration} {BICEP2, Keck Array}),\
  }\href {\doibase 10.1103/PhysRevLett.121.221301} {\bibfield  {journal}
  {\bibinfo  {journal} {Phys. Rev. Lett.}\ }\textbf {\bibinfo {volume} {121}},\
  \bibinfo {pages} {221301} (\bibinfo {year} {2018})},\ \Eprint
  {http://arxiv.org/abs/1810.05216} {arXiv:1810.05216 [astro-ph.CO]}
  \BibitemShut {NoStop}%
\bibitem [{\citenamefont {Battistelli}\ \emph {et~al.}(2011)\citenamefont
  {Battistelli} \emph {et~al.}}]{Battistelli:2010aa}%
  \BibitemOpen
  \bibfield  {author} {\bibinfo {author} {\bibfnamefont {E.}~\bibnamefont
  {Battistelli}} \emph {et~al.} (\bibinfo {collaboration} {QUBIC}),\ }\href
  {\doibase 10.1016/j.astropartphys.2011.01.012} {\bibfield  {journal}
  {\bibinfo  {journal} {Astropart. Phys.}\ }\textbf {\bibinfo {volume} {34}},\
  \bibinfo {pages} {705} (\bibinfo {year} {2011})},\ \Eprint
  {http://arxiv.org/abs/1010.0645} {arXiv:1010.0645 [astro-ph.IM]} \BibitemShut
  {NoStop}%
\bibitem [{\citenamefont {Battistelli}\ \emph {et~al.}(2020)\citenamefont
  {Battistelli} \emph {et~al.}}]{Battistelli:2020siw}%
  \BibitemOpen
  \bibfield  {author} {\bibinfo {author} {\bibfnamefont {E.}~\bibnamefont
  {Battistelli}} \emph {et~al.} (\bibinfo {collaboration} {QUBIC}),\ }\href
  {\doibase 10.1007/s10909-020-02370-0} {\bibfield  {journal} {\bibinfo
  {journal} {J. Low. Temp. Phys.}\ }\textbf {\bibinfo {volume} {199}},\
  \bibinfo {pages} {482} (\bibinfo {year} {2020})},\ \Eprint
  {http://arxiv.org/abs/2001.10272} {arXiv:2001.10272 [astro-ph.IM]}
  \BibitemShut {NoStop}%
\bibitem [{\citenamefont {Lyth}(1984)}]{Lyth:1984yz}%
  \BibitemOpen
  \bibfield  {author} {\bibinfo {author} {\bibfnamefont {D.}~\bibnamefont
  {Lyth}},\ }\href {\doibase 10.1016/0370-2693(84)91391-1} {\bibfield
  {journal} {\bibinfo  {journal} {Phys. Lett. B}\ }\textbf {\bibinfo {volume}
  {147}},\ \bibinfo {pages} {403} (\bibinfo {year} {1984})},\ \bibinfo {note}
  {[Erratum: Phys.Lett.B 150, 465 (1985)]}\BibitemShut {NoStop}%
\bibitem [{\citenamefont {Lyth}(1997)}]{Lyth:1996im}%
  \BibitemOpen
  \bibfield  {author} {\bibinfo {author} {\bibfnamefont {D.~H.}\ \bibnamefont
  {Lyth}},\ }\href {\doibase 10.1103/PhysRevLett.78.1861} {\bibfield  {journal}
  {\bibinfo  {journal} {Phys. Rev. Lett.}\ }\textbf {\bibinfo {volume} {78}},\
  \bibinfo {pages} {1861} (\bibinfo {year} {1997})},\ \Eprint
  {http://arxiv.org/abs/hep-ph/9606387} {arXiv:hep-ph/9606387} \BibitemShut
  {NoStop}%
\bibitem [{\citenamefont {Freese}\ \emph {et~al.}(1990)\citenamefont {Freese},
  \citenamefont {Frieman},\ and\ \citenamefont {Olinto}}]{Freese:1990rb}%
  \BibitemOpen
  \bibfield  {author} {\bibinfo {author} {\bibfnamefont {K.}~\bibnamefont
  {Freese}}, \bibinfo {author} {\bibfnamefont {J.~A.}\ \bibnamefont {Frieman}},
  \ and\ \bibinfo {author} {\bibfnamefont {A.~V.}\ \bibnamefont {Olinto}},\
  }\href {\doibase 10.1103/PhysRevLett.65.3233} {\bibfield  {journal} {\bibinfo
   {journal} {Phys. Rev. Lett.}\ }\textbf {\bibinfo {volume} {65}},\ \bibinfo
  {pages} {3233} (\bibinfo {year} {1990})}\BibitemShut {NoStop}%
\bibitem [{\citenamefont {Adams}\ \emph {et~al.}(1993)\citenamefont {Adams},
  \citenamefont {Bond}, \citenamefont {Freese}, \citenamefont {Frieman},\ and\
  \citenamefont {Olinto}}]{Adams:1992bn}%
  \BibitemOpen
  \bibfield  {author} {\bibinfo {author} {\bibfnamefont {F.~C.}\ \bibnamefont
  {Adams}}, \bibinfo {author} {\bibfnamefont {J.}~\bibnamefont {Bond}},
  \bibinfo {author} {\bibfnamefont {K.}~\bibnamefont {Freese}}, \bibinfo
  {author} {\bibfnamefont {J.~A.}\ \bibnamefont {Frieman}}, \ and\ \bibinfo
  {author} {\bibfnamefont {A.~V.}\ \bibnamefont {Olinto}},\ }\href {\doibase
  10.1103/PhysRevD.47.426} {\bibfield  {journal} {\bibinfo  {journal} {Phys.
  Rev. D}\ }\textbf {\bibinfo {volume} {47}},\ \bibinfo {pages} {426} (\bibinfo
  {year} {1993})},\ \Eprint {http://arxiv.org/abs/hep-ph/9207245}
  {arXiv:hep-ph/9207245} \BibitemShut {NoStop}%
\bibitem [{\citenamefont {Li}\ \emph {et~al.}(2015)\citenamefont {Li},
  \citenamefont {Li},\ and\ \citenamefont {Nanopoulos}}]{Li:2014vpa}%
  \BibitemOpen
  \bibfield  {author} {\bibinfo {author} {\bibfnamefont {T.}~\bibnamefont
  {Li}}, \bibinfo {author} {\bibfnamefont {Z.}~\bibnamefont {Li}}, \ and\
  \bibinfo {author} {\bibfnamefont {D.~V.}\ \bibnamefont {Nanopoulos}},\ }\href
  {\doibase 10.1103/PhysRevD.91.061303} {\bibfield  {journal} {\bibinfo
  {journal} {Phys. Rev. D}\ }\textbf {\bibinfo {volume} {91}},\ \bibinfo
  {pages} {061303} (\bibinfo {year} {2015})},\ \Eprint
  {http://arxiv.org/abs/1409.3267} {arXiv:1409.3267 [hep-th]} \BibitemShut
  {NoStop}%
\bibitem [{\citenamefont {McDonald}(2014)}]{McDonald:2014oza}%
  \BibitemOpen
  \bibfield  {author} {\bibinfo {author} {\bibfnamefont {J.}~\bibnamefont
  {McDonald}},\ }\href {\doibase 10.1088/1475-7516/2014/09/027} {\bibfield
  {journal} {\bibinfo  {journal} {JCAP}\ }\textbf {\bibinfo {volume} {09}},\
  \bibinfo {pages} {027} (\bibinfo {year} {2014})},\ \Eprint
  {http://arxiv.org/abs/1404.4620} {arXiv:1404.4620 [hep-ph]} \BibitemShut
  {NoStop}%
\bibitem [{\citenamefont {McDonald}(2015)}]{McDonald:2014nqa}%
  \BibitemOpen
  \bibfield  {author} {\bibinfo {author} {\bibfnamefont {J.}~\bibnamefont
  {McDonald}},\ }\href {\doibase 10.1088/1475-7516/2015/01/018} {\bibfield
  {journal} {\bibinfo  {journal} {JCAP}\ }\textbf {\bibinfo {volume} {01}},\
  \bibinfo {pages} {018} (\bibinfo {year} {2015})},\ \Eprint
  {http://arxiv.org/abs/1407.7471} {arXiv:1407.7471 [hep-ph]} \BibitemShut
  {NoStop}%
\bibitem [{\citenamefont {Ben-Dayan}\ and\ \citenamefont
  {Brustein}(2010)}]{BenDayan:2009kv}%
  \BibitemOpen
  \bibfield  {author} {\bibinfo {author} {\bibfnamefont {I.}~\bibnamefont
  {Ben-Dayan}}\ and\ \bibinfo {author} {\bibfnamefont {R.}~\bibnamefont
  {Brustein}},\ }\href {\doibase 10.1088/1475-7516/2010/09/007} {\bibfield
  {journal} {\bibinfo  {journal} {JCAP}\ }\textbf {\bibinfo {volume} {09}},\
  \bibinfo {pages} {007} (\bibinfo {year} {2010})},\ \Eprint
  {http://arxiv.org/abs/0907.2384} {arXiv:0907.2384 [astro-ph.CO]} \BibitemShut
  {NoStop}%
\bibitem [{\citenamefont {Choudhury}\ and\ \citenamefont
  {Mazumdar}(2014{\natexlab{a}})}]{Choudhury:2013iaa}%
  \BibitemOpen
  \bibfield  {author} {\bibinfo {author} {\bibfnamefont {S.}~\bibnamefont
  {Choudhury}}\ and\ \bibinfo {author} {\bibfnamefont {A.}~\bibnamefont
  {Mazumdar}},\ }\href {\doibase 10.1016/j.nuclphysb.2014.03.005} {\bibfield
  {journal} {\bibinfo  {journal} {Nucl. Phys. B}\ }\textbf {\bibinfo {volume}
  {882}},\ \bibinfo {pages} {386} (\bibinfo {year} {2014}{\natexlab{a}})},\
  \Eprint {http://arxiv.org/abs/1306.4496} {arXiv:1306.4496 [hep-ph]}
  \BibitemShut {NoStop}%
\bibitem [{\citenamefont {Choudhury}\ and\ \citenamefont
  {Mazumdar}(2014{\natexlab{b}})}]{Choudhury:2014kma}%
  \BibitemOpen
  \bibfield  {author} {\bibinfo {author} {\bibfnamefont {S.}~\bibnamefont
  {Choudhury}}\ and\ \bibinfo {author} {\bibfnamefont {A.}~\bibnamefont
  {Mazumdar}},\ }\href@noop {} {\  (\bibinfo {year} {2014}{\natexlab{b}})},\
  \Eprint {http://arxiv.org/abs/1403.5549} {arXiv:1403.5549 [hep-th]}
  \BibitemShut {NoStop}%
\bibitem [{\citenamefont {Antusch}\ and\ \citenamefont
  {Nolde}(2014)}]{Antusch:2014cpa}%
  \BibitemOpen
  \bibfield  {author} {\bibinfo {author} {\bibfnamefont {S.}~\bibnamefont
  {Antusch}}\ and\ \bibinfo {author} {\bibfnamefont {D.}~\bibnamefont
  {Nolde}},\ }\href {\doibase 10.1088/1475-7516/2014/05/035} {\bibfield
  {journal} {\bibinfo  {journal} {JCAP}\ }\textbf {\bibinfo {volume} {05}},\
  \bibinfo {pages} {035} (\bibinfo {year} {2014})},\ \Eprint
  {http://arxiv.org/abs/1404.1821} {arXiv:1404.1821 [hep-ph]} \BibitemShut
  {NoStop}%
\bibitem [{\citenamefont {Gao}\ \emph {et~al.}(2015)\citenamefont {Gao},
  \citenamefont {Gong},\ and\ \citenamefont {Li}}]{Gao:2014pca}%
  \BibitemOpen
  \bibfield  {author} {\bibinfo {author} {\bibfnamefont {Q.}~\bibnamefont
  {Gao}}, \bibinfo {author} {\bibfnamefont {Y.}~\bibnamefont {Gong}}, \ and\
  \bibinfo {author} {\bibfnamefont {T.}~\bibnamefont {Li}},\ }\href {\doibase
  10.1103/PhysRevD.91.063509} {\bibfield  {journal} {\bibinfo  {journal} {Phys.
  Rev. D}\ }\textbf {\bibinfo {volume} {91}},\ \bibinfo {pages} {063509}
  (\bibinfo {year} {2015})},\ \Eprint {http://arxiv.org/abs/1405.6451}
  {arXiv:1405.6451 [gr-qc]} \BibitemShut {NoStop}%
\bibitem [{\citenamefont {Efstathiou}\ and\ \citenamefont
  {Mack}(2005)}]{Efstathiou:2005tq}%
  \BibitemOpen
  \bibfield  {author} {\bibinfo {author} {\bibfnamefont {G.}~\bibnamefont
  {Efstathiou}}\ and\ \bibinfo {author} {\bibfnamefont {K.~J.}\ \bibnamefont
  {Mack}},\ }\href {\doibase 10.1088/1475-7516/2005/05/008} {\bibfield
  {journal} {\bibinfo  {journal} {JCAP}\ }\textbf {\bibinfo {volume} {05}},\
  \bibinfo {pages} {008} (\bibinfo {year} {2005})},\ \Eprint
  {http://arxiv.org/abs/astro-ph/0503360} {arXiv:astro-ph/0503360} \BibitemShut
  {NoStop}%
\bibitem [{\citenamefont {Baumann}\ and\ \citenamefont
  {Green}(2012)}]{Baumann:2011ws}%
  \BibitemOpen
  \bibfield  {author} {\bibinfo {author} {\bibfnamefont {D.}~\bibnamefont
  {Baumann}}\ and\ \bibinfo {author} {\bibfnamefont {D.}~\bibnamefont
  {Green}},\ }\href {\doibase 10.1088/1475-7516/2012/05/017} {\bibfield
  {journal} {\bibinfo  {journal} {JCAP}\ }\textbf {\bibinfo {volume} {05}},\
  \bibinfo {pages} {017} (\bibinfo {year} {2012})},\ \Eprint
  {http://arxiv.org/abs/1111.3040} {arXiv:1111.3040 [hep-th]} \BibitemShut
  {NoStop}%
\bibitem [{\citenamefont {Garcia-Bellido}\ \emph
  {et~al.}(2014{\natexlab{a}})\citenamefont {Garcia-Bellido}, \citenamefont
  {Roest}, \citenamefont {Scalisi},\ and\ \citenamefont
  {Zavala}}]{Garcia-Bellido:2014eva}%
  \BibitemOpen
  \bibfield  {author} {\bibinfo {author} {\bibfnamefont {J.}~\bibnamefont
  {Garcia-Bellido}}, \bibinfo {author} {\bibfnamefont {D.}~\bibnamefont
  {Roest}}, \bibinfo {author} {\bibfnamefont {M.}~\bibnamefont {Scalisi}}, \
  and\ \bibinfo {author} {\bibfnamefont {I.}~\bibnamefont {Zavala}},\ }\href
  {\doibase 10.1088/1475-7516/2014/09/006} {\bibfield  {journal} {\bibinfo
  {journal} {JCAP}\ }\textbf {\bibinfo {volume} {09}},\ \bibinfo {pages} {006}
  (\bibinfo {year} {2014}{\natexlab{a}})},\ \Eprint
  {http://arxiv.org/abs/1405.7399} {arXiv:1405.7399 [hep-th]} \BibitemShut
  {NoStop}%
\bibitem [{\citenamefont {Garcia-Bellido}\ \emph
  {et~al.}(2014{\natexlab{b}})\citenamefont {Garcia-Bellido}, \citenamefont
  {Roest}, \citenamefont {Scalisi},\ and\ \citenamefont
  {Zavala}}]{Garcia-Bellido:2014wfa}%
  \BibitemOpen
  \bibfield  {author} {\bibinfo {author} {\bibfnamefont {J.}~\bibnamefont
  {Garcia-Bellido}}, \bibinfo {author} {\bibfnamefont {D.}~\bibnamefont
  {Roest}}, \bibinfo {author} {\bibfnamefont {M.}~\bibnamefont {Scalisi}}, \
  and\ \bibinfo {author} {\bibfnamefont {I.}~\bibnamefont {Zavala}},\ }\href
  {\doibase 10.1103/PhysRevD.90.123539} {\bibfield  {journal} {\bibinfo
  {journal} {Phys. Rev. D}\ }\textbf {\bibinfo {volume} {90}},\ \bibinfo
  {pages} {123539} (\bibinfo {year} {2014}{\natexlab{b}})},\ \Eprint
  {http://arxiv.org/abs/1408.6839} {arXiv:1408.6839 [hep-th]} \BibitemShut
  {NoStop}%
\bibitem [{\citenamefont {Huang}(2015)}]{Huang:2015xda}%
  \BibitemOpen
  \bibfield  {author} {\bibinfo {author} {\bibfnamefont {Q.-G.}\ \bibnamefont
  {Huang}},\ }\href {\doibase 10.1103/PhysRevD.91.123532} {\bibfield  {journal}
  {\bibinfo  {journal} {Phys. Rev. D}\ }\textbf {\bibinfo {volume} {91}},\
  \bibinfo {pages} {123532} (\bibinfo {year} {2015})},\ \Eprint
  {http://arxiv.org/abs/1503.04513} {arXiv:1503.04513 [astro-ph.CO]}
  \BibitemShut {NoStop}%
\bibitem [{\citenamefont {Linde}(2017)}]{Linde:2016hbb}%
  \BibitemOpen
  \bibfield  {author} {\bibinfo {author} {\bibfnamefont {A.}~\bibnamefont
  {Linde}},\ }\href {\doibase 10.1088/1475-7516/2017/02/006} {\bibfield
  {journal} {\bibinfo  {journal} {JCAP}\ }\textbf {\bibinfo {volume} {02}},\
  \bibinfo {pages} {006} (\bibinfo {year} {2017})},\ \Eprint
  {http://arxiv.org/abs/1612.00020} {arXiv:1612.00020 [astro-ph.CO]}
  \BibitemShut {NoStop}%
\bibitem [{\citenamefont {Di~Marco}(2017)}]{DiMarco:2017ihz}%
  \BibitemOpen
  \bibfield  {author} {\bibinfo {author} {\bibfnamefont {A.}~\bibnamefont
  {Di~Marco}},\ }\href {\doibase 10.1103/PhysRevD.96.023511} {\bibfield
  {journal} {\bibinfo  {journal} {Phys. Rev. D}\ }\textbf {\bibinfo {volume}
  {96}},\ \bibinfo {pages} {023511} (\bibinfo {year} {2017})},\ \Eprint
  {http://arxiv.org/abs/1706.04144} {arXiv:1706.04144 [astro-ph.CO]}
  \BibitemShut {NoStop}%
\bibitem [{\citenamefont {Lyth}\ and\ \citenamefont
  {Riotto}(1999)}]{Lyth:1998xn}%
  \BibitemOpen
  \bibfield  {author} {\bibinfo {author} {\bibfnamefont {D.~H.}\ \bibnamefont
  {Lyth}}\ and\ \bibinfo {author} {\bibfnamefont {A.}~\bibnamefont {Riotto}},\
  }\href {\doibase 10.1016/S0370-1573(98)00128-8} {\bibfield  {journal}
  {\bibinfo  {journal} {Phys. Rept.}\ }\textbf {\bibinfo {volume} {314}},\
  \bibinfo {pages} {1} (\bibinfo {year} {1999})},\ \Eprint
  {http://arxiv.org/abs/hep-ph/9807278} {arXiv:hep-ph/9807278} \BibitemShut
  {NoStop}%
\bibitem [{\citenamefont {Tsamis}\ and\ \citenamefont
  {Woodard}(2004)}]{PhysRevD.69.084005}%
  \BibitemOpen
  \bibfield  {author} {\bibinfo {author} {\bibfnamefont {N.~C.}\ \bibnamefont
  {Tsamis}}\ and\ \bibinfo {author} {\bibfnamefont {R.~P.}\ \bibnamefont
  {Woodard}},\ }\href {\doibase 10.1103/PhysRevD.69.084005} {\bibfield
  {journal} {\bibinfo  {journal} {Phys. Rev. D}\ }\textbf {\bibinfo {volume}
  {69}},\ \bibinfo {pages} {084005} (\bibinfo {year} {2004})}\BibitemShut
  {NoStop}%
\bibitem [{\citenamefont {Kinney}(2005)}]{PhysRevD.72.023515}%
  \BibitemOpen
  \bibfield  {author} {\bibinfo {author} {\bibfnamefont {W.~H.}\ \bibnamefont
  {Kinney}},\ }\href {\doibase 10.1103/PhysRevD.72.023515} {\bibfield
  {journal} {\bibinfo  {journal} {Phys. Rev. D}\ }\textbf {\bibinfo {volume}
  {72}},\ \bibinfo {pages} {023515} (\bibinfo {year} {2005})}\BibitemShut
  {NoStop}%
\bibitem [{\citenamefont {{Hirano}}\ \emph {et~al.}(2016)\citenamefont
  {{Hirano}}, \citenamefont {{Kobayashi}},\ and\ \citenamefont
  {{Yokoyama}}}]{2016PhRvD..94j3515H}%
  \BibitemOpen
  \bibfield  {author} {\bibinfo {author} {\bibfnamefont {S.}~\bibnamefont
  {{Hirano}}}, \bibinfo {author} {\bibfnamefont {T.}~\bibnamefont
  {{Kobayashi}}}, \ and\ \bibinfo {author} {\bibfnamefont {S.}~\bibnamefont
  {{Yokoyama}}},\ }\href {\doibase 10.1103/PhysRevD.94.103515} {\bibfield
  {journal} {\bibinfo  {journal} {\prd}\ }\textbf {\bibinfo {volume} {94}},\
  \bibinfo {eid} {103515} (\bibinfo {year} {2016})},\ \Eprint
  {http://arxiv.org/abs/1604.00141} {arXiv:1604.00141 [astro-ph.CO]}
  \BibitemShut {NoStop}%
\bibitem [{\citenamefont {Dimopoulos}(2017)}]{DIMOPOULOS2017262}%
  \BibitemOpen
  \bibfield  {author} {\bibinfo {author} {\bibfnamefont {K.}~\bibnamefont
  {Dimopoulos}},\ }\href {\doibase
  https://doi.org/10.1016/j.physletb.2017.10.066} {\bibfield  {journal}
  {\bibinfo  {journal} {Physics Letters B}\ }\textbf {\bibinfo {volume}
  {775}},\ \bibinfo {pages} {262 } (\bibinfo {year} {2017})}\BibitemShut
  {NoStop}%
\bibitem [{\citenamefont {Easther}\ \emph {et~al.}(2006)\citenamefont
  {Easther}, \citenamefont {Kinney},\ and\ \citenamefont
  {Powell}}]{Easther:2006qu}%
  \BibitemOpen
  \bibfield  {author} {\bibinfo {author} {\bibfnamefont {R.}~\bibnamefont
  {Easther}}, \bibinfo {author} {\bibfnamefont {W.~H.}\ \bibnamefont {Kinney}},
  \ and\ \bibinfo {author} {\bibfnamefont {B.~A.}\ \bibnamefont {Powell}},\
  }\href {\doibase 10.1088/1475-7516/2006/08/004} {\bibfield  {journal}
  {\bibinfo  {journal} {JCAP}\ }\textbf {\bibinfo {volume} {08}},\ \bibinfo
  {pages} {004} (\bibinfo {year} {2006})},\ \Eprint
  {http://arxiv.org/abs/astro-ph/0601276} {arXiv:astro-ph/0601276} \BibitemShut
  {NoStop}%
\bibitem [{\citenamefont {Sasaki}(1986)}]{10.1143/PTP.76.1036}%
  \BibitemOpen
  \bibfield  {author} {\bibinfo {author} {\bibfnamefont {M.}~\bibnamefont
  {Sasaki}},\ }\href {\doibase 10.1143/PTP.76.1036} {\bibfield  {journal}
  {\bibinfo  {journal} {Progress of Theoretical Physics}\ }\textbf {\bibinfo
  {volume} {76}},\ \bibinfo {pages} {1036} (\bibinfo {year} {1986})},\ \Eprint
  {http://arxiv.org/abs/https://academic.oup.com/ptp/article-pdf/76/5/1036/5152623/76-5-1036.pdf}
  {https://academic.oup.com/ptp/article-pdf/76/5/1036/5152623/76-5-1036.pdf}
  \BibitemShut {NoStop}%
\bibitem [{\citenamefont {Mukhanov}(1988)}]{Mukhanov:1988jd}%
  \BibitemOpen
  \bibfield  {author} {\bibinfo {author} {\bibfnamefont {V.~F.}\ \bibnamefont
  {Mukhanov}},\ }\href@noop {} {\bibfield  {journal} {\bibinfo  {journal} {Sov.
  Phys. JETP}\ }\textbf {\bibinfo {volume} {67}},\ \bibinfo {pages} {1297}
  (\bibinfo {year} {1988})}\BibitemShut {NoStop}%
\bibitem [{\citenamefont {Stewart}\ and\ \citenamefont
  {Lyth}(1993)}]{STEWART1993171}%
  \BibitemOpen
  \bibfield  {author} {\bibinfo {author} {\bibfnamefont {E.~D.}\ \bibnamefont
  {Stewart}}\ and\ \bibinfo {author} {\bibfnamefont {D.~H.}\ \bibnamefont
  {Lyth}},\ }\href {\doibase https://doi.org/10.1016/0370-2693(93)90379-V}
  {\bibfield  {journal} {\bibinfo  {journal} {Physics Letters B}\ }\textbf
  {\bibinfo {volume} {302}},\ \bibinfo {pages} {171 } (\bibinfo {year}
  {1993})}\BibitemShut {NoStop}%
\bibitem [{\citenamefont {Casadio}\ \emph {et~al.}(2005)\citenamefont
  {Casadio}, \citenamefont {Finelli}, \citenamefont {Luzzi},\ and\
  \citenamefont {Venturi}}]{CASADIO20051}%
  \BibitemOpen
  \bibfield  {author} {\bibinfo {author} {\bibfnamefont {R.}~\bibnamefont
  {Casadio}}, \bibinfo {author} {\bibfnamefont {F.}~\bibnamefont {Finelli}},
  \bibinfo {author} {\bibfnamefont {M.}~\bibnamefont {Luzzi}}, \ and\ \bibinfo
  {author} {\bibfnamefont {G.}~\bibnamefont {Venturi}},\ }\href {\doibase
  https://doi.org/10.1016/j.physletb.2005.08.056} {\bibfield  {journal}
  {\bibinfo  {journal} {Physics Letters B}\ }\textbf {\bibinfo {volume}
  {625}},\ \bibinfo {pages} {1 } (\bibinfo {year} {2005})}\BibitemShut
  {NoStop}%
\bibitem [{\citenamefont {Kogut}\ \emph {et~al.}(2011)\citenamefont {Kogut}
  \emph {et~al.}}]{Kogut:2011xw}%
  \BibitemOpen
  \bibfield  {author} {\bibinfo {author} {\bibfnamefont {A.}~\bibnamefont
  {Kogut}} \emph {et~al.},\ }\href {\doibase 10.1088/1475-7516/2011/07/025}
  {\bibfield  {journal} {\bibinfo  {journal} {JCAP}\ }\textbf {\bibinfo
  {volume} {07}},\ \bibinfo {pages} {025} (\bibinfo {year} {2011})},\ \Eprint
  {http://arxiv.org/abs/1105.2044} {arXiv:1105.2044 [astro-ph.CO]} \BibitemShut
  {NoStop}%
\bibitem [{\citenamefont {Bouchet}\ \emph {et~al.}(2011)\citenamefont {Bouchet}
  \emph {et~al.}}]{Bouchet:2011ck}%
  \BibitemOpen
  \bibfield  {author} {\bibinfo {author} {\bibfnamefont {F.}~\bibnamefont
  {Bouchet}} \emph {et~al.} (\bibinfo {collaboration} {COrE}),\ }\href@noop {}
  {\  (\bibinfo {year} {2011})},\ \Eprint {http://arxiv.org/abs/1102.2181}
  {arXiv:1102.2181 [astro-ph.CO]} \BibitemShut {NoStop}%
\bibitem [{\citenamefont {Matsumura}\ \emph {et~al.}(2014)\citenamefont
  {Matsumura} \emph {et~al.}}]{Matsumura:2013aja}%
  \BibitemOpen
  \bibfield  {author} {\bibinfo {author} {\bibfnamefont {T.}~\bibnamefont
  {Matsumura}} \emph {et~al.},\ }\href {\doibase 10.1007/s10909-013-0996-1}
  {\bibfield  {journal} {\bibinfo  {journal} {J. Low Temp. Phys.}\ }\textbf
  {\bibinfo {volume} {176}},\ \bibinfo {pages} {733} (\bibinfo {year}
  {2014})},\ \Eprint {http://arxiv.org/abs/1311.2847} {arXiv:1311.2847
  [astro-ph.IM]} \BibitemShut {NoStop}%
\bibitem [{\citenamefont {Lazear}\ \emph {et~al.}(2014)\citenamefont {Lazear},
  \citenamefont {Ade}, \citenamefont {Benford}, \citenamefont {Bennett},
  \citenamefont {Chuss}, \citenamefont {Dotson}, \citenamefont {Eimer},
  \citenamefont {Fixsen}, \citenamefont {Halpern}, \citenamefont {Hilton} \emph
  {et~al.}}]{lazear2014primordial}%
  \BibitemOpen
  \bibfield  {author} {\bibinfo {author} {\bibfnamefont {J.}~\bibnamefont
  {Lazear}}, \bibinfo {author} {\bibfnamefont {P.~A.}\ \bibnamefont {Ade}},
  \bibinfo {author} {\bibfnamefont {D.}~\bibnamefont {Benford}}, \bibinfo
  {author} {\bibfnamefont {C.~L.}\ \bibnamefont {Bennett}}, \bibinfo {author}
  {\bibfnamefont {D.~T.}\ \bibnamefont {Chuss}}, \bibinfo {author}
  {\bibfnamefont {J.~L.}\ \bibnamefont {Dotson}}, \bibinfo {author}
  {\bibfnamefont {J.~R.}\ \bibnamefont {Eimer}}, \bibinfo {author}
  {\bibfnamefont {D.~J.}\ \bibnamefont {Fixsen}}, \bibinfo {author}
  {\bibfnamefont {M.}~\bibnamefont {Halpern}}, \bibinfo {author} {\bibfnamefont
  {G.}~\bibnamefont {Hilton}},  \emph {et~al.},\ }\bibfield  {booktitle} {\emph
  {\bibinfo {booktitle} {Millimeter, Submillimeter, and Far-Infrared Detectors
  and Instrumentation for Astronomy VII}},\ }\href@noop {} {\ \textbf {\bibinfo
  {volume} {9153}},\ \bibinfo {pages} {91531L} (\bibinfo {year}
  {2014})}\BibitemShut {NoStop}%
\bibitem [{\citenamefont {Essinger-Hileman}\ \emph {et~al.}(2014)\citenamefont
  {Essinger-Hileman}, \citenamefont {Ali}, \citenamefont {Amiri}, \citenamefont
  {Appel}, \citenamefont {Araujo}, \citenamefont {Bennett}, \citenamefont
  {Boone}, \citenamefont {Chan}, \citenamefont {Cho}, \citenamefont {Chuss}
  \emph {et~al.}}]{essinger2014class}%
  \BibitemOpen
  \bibfield  {author} {\bibinfo {author} {\bibfnamefont {T.}~\bibnamefont
  {Essinger-Hileman}}, \bibinfo {author} {\bibfnamefont {A.}~\bibnamefont
  {Ali}}, \bibinfo {author} {\bibfnamefont {M.}~\bibnamefont {Amiri}}, \bibinfo
  {author} {\bibfnamefont {J.~W.}\ \bibnamefont {Appel}}, \bibinfo {author}
  {\bibfnamefont {D.}~\bibnamefont {Araujo}}, \bibinfo {author} {\bibfnamefont
  {C.~L.}\ \bibnamefont {Bennett}}, \bibinfo {author} {\bibfnamefont
  {F.}~\bibnamefont {Boone}}, \bibinfo {author} {\bibfnamefont
  {M.}~\bibnamefont {Chan}}, \bibinfo {author} {\bibfnamefont {H.-M.}\
  \bibnamefont {Cho}}, \bibinfo {author} {\bibfnamefont {D.~T.}\ \bibnamefont
  {Chuss}},  \emph {et~al.},\ }\bibfield  {booktitle} {\emph {\bibinfo
  {booktitle} {Millimeter, Submillimeter, and Far-Infrared Detectors and
  Instrumentation for Astronomy VII}},\ }\href@noop {} {\ \textbf {\bibinfo
  {volume} {9153}},\ \bibinfo {pages} {91531I} (\bibinfo {year}
  {2014})}\BibitemShut {NoStop}%
\bibitem [{\citenamefont {Finelli}\ \emph {et~al.}(2018)\citenamefont {Finelli}
  \emph {et~al.}}]{Finelli:2016cyd}%
  \BibitemOpen
  \bibfield  {author} {\bibinfo {author} {\bibfnamefont {F.}~\bibnamefont
  {Finelli}} \emph {et~al.} (\bibinfo {collaboration} {CORE}),\ }\href
  {\doibase 10.1088/1475-7516/2018/04/016} {\bibfield  {journal} {\bibinfo
  {journal} {JCAP}\ }\textbf {\bibinfo {volume} {04}},\ \bibinfo {pages} {016}
  (\bibinfo {year} {2018})},\ \Eprint {http://arxiv.org/abs/1612.08270}
  {arXiv:1612.08270 [astro-ph.CO]} \BibitemShut {NoStop}%
\bibitem [{\citenamefont {Crowder}\ and\ \citenamefont
  {Cornish}(2005)}]{Crowder:2005nr}%
  \BibitemOpen
  \bibfield  {author} {\bibinfo {author} {\bibfnamefont {J.}~\bibnamefont
  {Crowder}}\ and\ \bibinfo {author} {\bibfnamefont {N.~J.}\ \bibnamefont
  {Cornish}},\ }\href {\doibase 10.1103/PhysRevD.72.083005} {\bibfield
  {journal} {\bibinfo  {journal} {Phys. Rev. D}\ }\textbf {\bibinfo {volume}
  {72}},\ \bibinfo {pages} {083005} (\bibinfo {year} {2005})},\ \Eprint
  {http://arxiv.org/abs/gr-qc/0506015} {arXiv:gr-qc/0506015} \BibitemShut
  {NoStop}%
\end{thebibliography}%
\end{document}